# Electric-field-induced domain walls in wurtzite ferroelectrics

*Ding Wang[1]†, Danhao Wang[1]*†, Mahlet Molla[2]†, Yujie Liu[2], Samuel Yang[1], Mingtao Hu[1], Jiangnan Liu[1], Yuanpeng Wu[1], Tao Ma[3], Emmanouil Kioupakis[2]*, Zetian Mi[1]**

[1]Department of Electrical Engineering and Computer Science, University of Michigan, Ann Arbor, 48109, USA.

[2]Department of Material Science and Engineering, University of Michigan, Ann Arbor, 48109, USA.

[3]Michigan Center for Materials Characterization, University of Michigan, Ann Arbor, 48109, USA.

*Corresponding author. Email: danhaow@umich.edu; kioup@umich.edu; ztmi@umich.edu

† These authors contributed equally to this work.

**Wurtzite ferroelectrics possess transformative potential for next-generation microelectronics. A comprehensive understanding of their ferroelectric properties and domain energetics is crucial for tailoring their ferroelectric characteristics and exploiting their functional properties in practical devices. Despite burgeoning interest, the exact configurations, and electronic structures of the domain walls in wurtzite ferroelectrics remain elusive. In this work, we elucidate the atomic configurations and electronic properties of electric-field-induced domain walls in ferroelectric ScGaN. By combining transmission electron microscopy and theoretical calculations, a novel charged domain wall with a buckled two-dimensional hexagonal phase is revealed. The dangling bonds associated with these domain walls give rise to unprecedented metallic-like mid-gap states within the forbidden band. Quantitative analysis further unveils a universal charge-compensation mechanism stabilizing antipolar domain walls in ferroelectric materials, wherein the**

**polarization discontinuity at the 180º domain wall is compensated by the dangling bond electrons. Furthermore, the reconfigurable conductivity of these domain walls is experimentally demonstrated, showcasing their potential for ultra-scaled device applications. Our findings represent a pivotal advancement in understanding the structural and electronic properties of wurtzite ferroelectric domain walls and lay the groundwork for fundamental physics studies and device applications.**

## Main Text

The recent discovery of ferroelectricity in wurtzite nitride semiconductors has sparked renewed interest in the electronics, piezoelectronics and photonics communities[1-3]. This new class of ferroelectrics, including ScAlN, BAlN, ScGaN, YAlN, and their alloys, possesses tantalizing properties, including low dielectric constants, scaling to nanometer dimensions, tunable coercive fields ($E_c$), high remanent polarizations ($P_r$), and thermal stability[4-15]. These characteristics position them as potential game-changers in modern microelectronics, from high-frequency resonators[10] to advanced memory[16] and computing architectures[17-19], all while being seamlessly integrable with mainstream semiconductor platforms[20]. Fully harnessing their capabilities and meeting the rigorous demands of modern electronics, however, necessitates refining critical parameters such as coercive field, endurance, stability, and leakage[1, 2]. This demands foundational studies beyond indirect macroscopic methods to unravel the underlying physics of ferroelectric interfaces at the unit cell level[21-25]. Recently, researchers have explored the transient polarity reversal in ferroelectric $Al_{0.94}B_{0.06}N$ triggered by electron beam exposure in transmission electron microscopy (TEM)[6]. However, the polarity switching mechanism under the influence of an external electric field, which is essential for the operation of any ferroelectric device and occurs on a much shorter time scale, may be entirely different[25].

A fundamental understanding of the intricate microstructures at the atomic scale is not only critical for addressing the contemporary challenges in wurtzite ferroelectrics, but also essential for exploiting emerging domain wall electronics[26]. The quasi-two-dimensional nature of domain walls, coupled with their distinct symmetry and local chemical composition, enable a spectrum of unconventional electric, magnetic, and optical properties, offering a promising path for next-generation nanoelectronics, such as domain-wall memory and ferroelectric field-effect transistors [27, 28]. In wurtzite semiconductors, the large spontaneous polarization and inherent defects, such as nitrogen vacancies and interstitials, further unfurl a wealth of design opportunities, facilitating the manipulation of crystallographic structures and domain wall electric, spin, and orbital interactions[26]. Despite these promising horizons, the exact atomic configurations and electronic structures of domain walls, especially the 180° domain wall in wurtzite ferroelectrics, remain poorly understood, hindering their practical applications in functional nanodevices.

In this study, we conduct a detailed investigation of the electric-field-induced domain walls in ScGaN, a representative wurtzite ferroelectric. The fully epitaxial and monocrystalline nature of ScGaN grown by molecular beam epitaxy (MBE) enables direct atomic-scale observations of localized polar order within the wurtzite lattice. Sub-nanometer vertical domain walls (VDWs) with antiparallel polarization and horizontal domain walls (HDWs) with antipolar configuration, both created via external electric fields, are unveiled by combining sub-angstrom resolution scanning transmission electron microscopy (STEM) and density functional theory (DFT) calculations. In particular, the electric-field-induced 180° HDWs, only a few monolayers in thickness and featuring a large number of metal dangling bonds, reveal a novel buckled 2D hexagonal nitride phase with unprecedented metallic-like mid-gap states. A quantitative comparison of the charges carried by the dangling bonds to the bound polarization charge further reveals a universal charge-compensation mechanism in wurtzite ferroelectrics. Building on these results, reconfigurable conductive domain walls in nitride ferroelectrics are experimentally demonstrated. Our results mark a pivotal advancement in understanding the switching mechanism and domain wall energetics in wurtzite ferroelectrics and opens new avenues for designing next-generation nitride domain wall-based nanodevices.

### Atomic-scale polarity switching in monocrystalline wurtzite ferroelectric

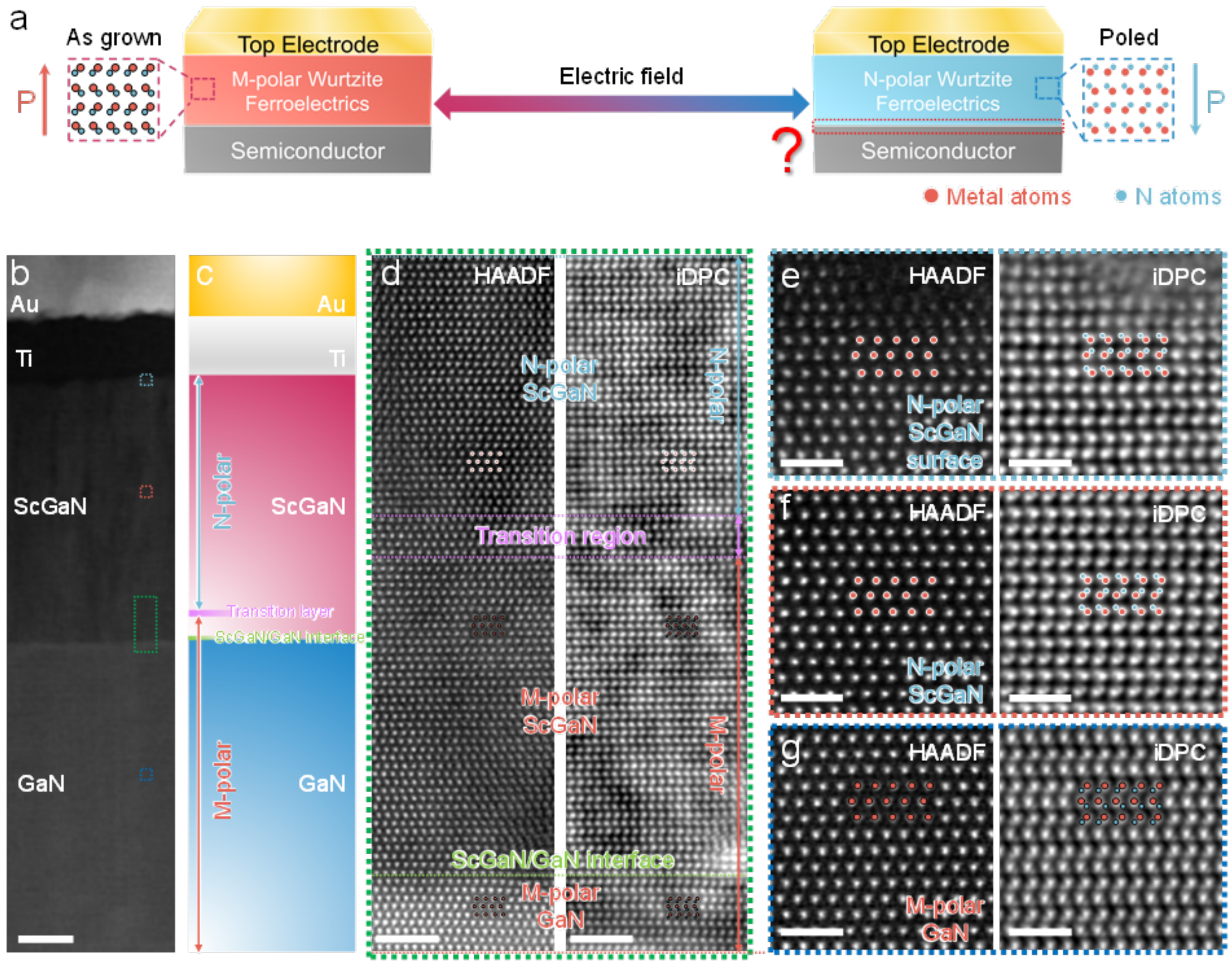

**Fig. 1. Atomic-scale evidence of polarity switching in ScGaN. a,** Schematic of polarity switching in a metal/ferroelectric/semiconductor capacitor. **b**, Cross-sectional TEM overview, and **c**, schematic illustration of the metal/ScGaN/GaN capacitor. **d**, High resolution HAADF-STEM

and iDPC images captured near the ScGaN/GaN interface. The atomic arrangement is represented by spheres (red for metal, blue for nitrogen). **e-g,** Magnified images from different regions. N-polar regions are clearly observed from above the transition region to the top surface of ScGaN. Scale bar: **b**, 20 nm; **d**, 2 nm; **e-g**, 1 nm.

**Fig. 1a** illustrates the concept of this work. The switchable ferroelectric and the unswitchable semiconductor electrode underpin a region of interest near the interface potentially rich in domain walls yet remains largely unexplored. ScGaN is compatible with mainstream semiconductor platforms and has a moderate bandgap, which is suitable for exploring the properties of wurtzite ferroelectrics[8]. The atomic-scale controllability of MBE ensures fully epitaxial layers with precise lattice alignment, atomically sharp interfaces and uniform elemental distribution (**Extended Data Fig. 1**), readily permitting atomic-level investigations of the interface and domain wall structures[29-31]. Prior to TEM studies, the ferroelectric nature of ScGaN is confirmed by a series of electrical and piezo-response force microscopy (PFM) measurements (**Extended Data Fig. 2**). We employ high-resolution STEM imaging with a high-angle annular dark-field (HAADF) detector and integrated differential phase contrast (iDPC) to visualize the metal and nitrogen atomic positions and determine crystallographic orientation. After electrical poling, different regions can be defined by their polarization orientations, schematically shown in **Fig. 1b, c**. Illustrated in the right panel of **Fig. 1d**, the alignment of atoms is clearly discernible, with Sc or Ga metal (M) and nitrogen (N) atoms positioned on opposing ends of the dumbbell-shaped atomic pairs. The polarity of the layer stack is subsequently ascertained from the stacking order of the M and N atoms within these dumbbells. **Fig. 1d** demonstrates the coexistence of electric-field-induced nitrogen-polar (N-polar) regions, an intermediary realm marked by mixed polarity, and unswitched metal-polar (M-polar) regions close to the M-polar GaN substrate (**Fig. 1g**). In contrast, the as-fabricated ScGaN/GaN heterostructure, without any prior poling, displays exclusively metal polarity (**Extended Data Fig.**

**1d**). The presence of N polarity in switched ScGaN films, extending from the transition region up to the top electrode (**Fig. 1d-f**), agrees well with the results of electrical and PFM characterizations, suggesting that the N-polar domains result from a comprehensive polarity reversal, rather than from localized defects or specific domain structures generated during growth. The observation of these distinct N-polar regions after electrical poling thus provides unambiguous evidence of ferroelectric polarization switching in epitaxial Sc-III-nitrides. We note that the existence of an unswitched layer at the ferroelectric/electrode interface has also been observed in ScAlN and other ferroelectric systems, likely due to a combination of domain energy and incomplete charge compensation[23, 32]. Here, this unswitched region is harnessed to promote the formation of in-plane domain walls; nonetheless, in practical applications, this unswitched layer will need to be tailored to enable device scale-down.

**Charge neutral vertical domain walls**

Domains and domain walls are not only fundamental to polarity switching dynamics but also represent energetically favorable configurations amidst significant depolarization fields[33]. In ferroelectrics with out-of-plane polarizations, vertical domain walls (VDWs) typically form to minimize the energy and area of the domain walls[34]. **Fig. 2a** illustrates such a "side-by-side" domain arrangement, where adjacent domains with antiparallel polarizations are separated by an abrupt polarity transition. The intrinsic crystal symmetry of the wurtzite material suggests a lateral polarity inversion through a unique 8- and 4-fold ring arrangement, as shown in **Fig. 2b**. Such VDW structures have previously been referred to as IDB* in conventional III-nitride epitaxial films[35, 36], yet their exact form and whether they can be electrically controlled remain unknown. **Fig. 2c-f** show high-resolution STEM images in the transition region of polarity reversed ScGaN, clearly showing the presence of distinct, atomically sharp VDWs. These domain walls are characterized by a notable step in each metal plane between the M-polar and N-polar regions (**Fig.**

**2c** and **Extended Data Fig. 3a**). At the domain wall, metal atoms from opposing polarities overlap slightly within one or two columns, forming a dimer-like pattern along the [0001] direction. To determine the precise position of each metal atom column, two-dimensional Gaussian fitting of the STEM images is conducted[37]. Notably, the in-plane atomic spacing appears moderately expanded at the VDW, as shown in the in-plane distance map (**Fig. 2d**). Integrated differential phase contrast (iDPC) images are examined to locate the position of N atoms. By collectively fitting the positions of N and M atoms, the angles of the M-N dumbbells are determined (**Fig. 2e, f**), delineating a clear polarity transition across the metal dimer-like region. The corresponding differentiated differential phase contrast (dDPC) images of the VDW structure further unveil a unique atomic arrangement comprising of two metal and two nitrogen atoms (**Fig. 2g**). The atomic contrast measurements enable precise determination of interatomic distance within this structure (**Fig. 2h**). We identify a slightly larger separation between metal atoms (0.123$c$), compared to nitrogen atoms (0.111$c$), along the [0001] direction, where $c$ is the lattice constant of ScGaN away from the interface (**Fig. 2i**).

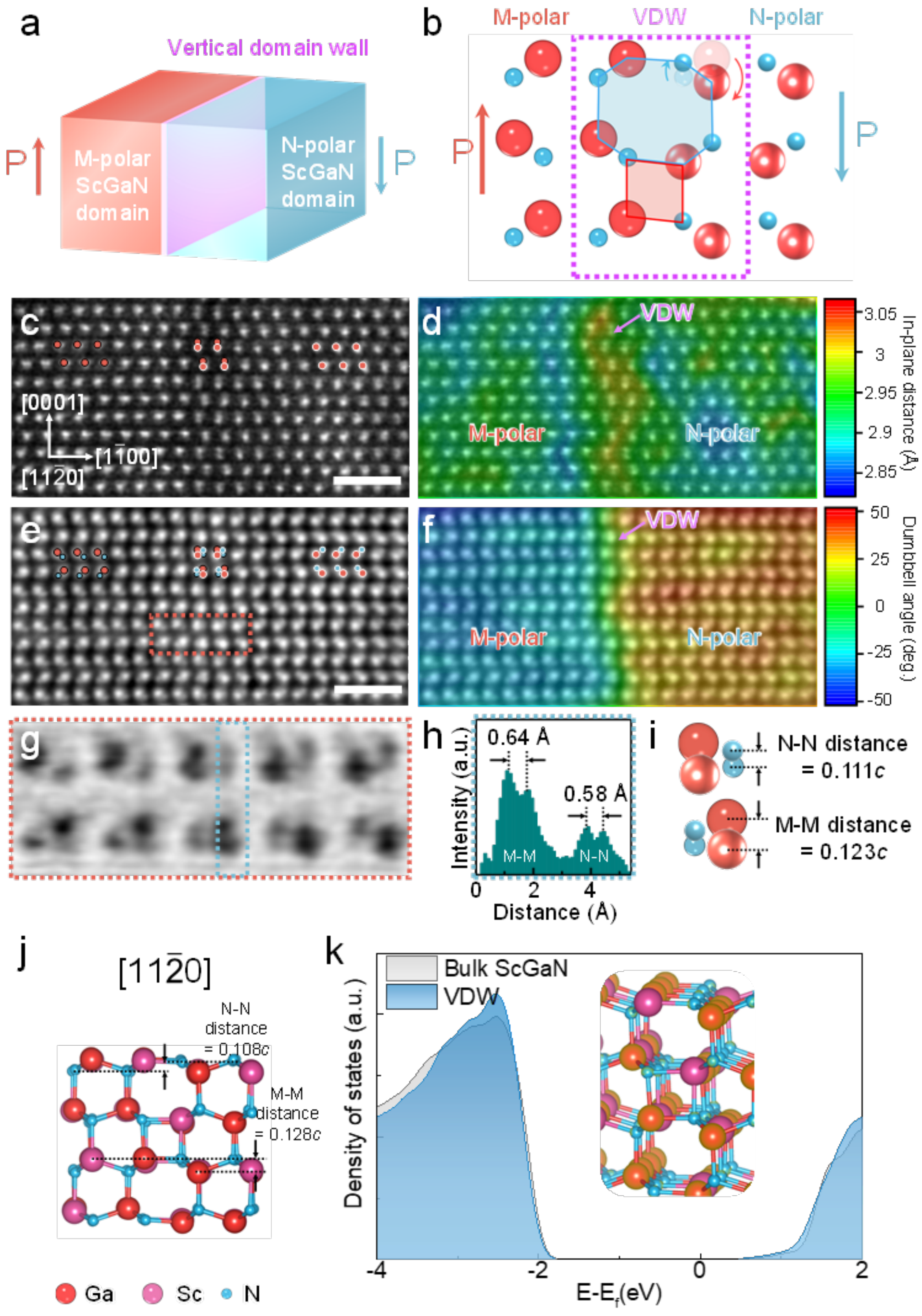

**Fig. 2. Electric-field-induced vertical domain walls in ScGaN. a,** Schematic, and **b,** atomic model of the vertical domain wall, with atoms represented by spheres (red for metal, blue for nitrogen). **c,** High magnification HAADF-STEM image and **d,** corresponding in-plane distance map of a vertical domain wall in ScGaN. **e,** High magnification iDPC-STEM images, and **f,** corresponding dumbbell angle map for the same region shown in **c** and **d**. Scale bar for **c-f**: 1 nm.

**g,** High-resolution dDPC image of the vertical domain wall structure. **h,** Intensity profile of the selected region in **g**. **i,** Measured average atomic distances in the dimer-like pattern in the vertical domain wall structure. **j,** DFT model of the vertical domain wall structure viewed along [$11\bar{2}0$]. The calculated average atomic distances are labeled in comparison with **i**. **k,** Electronic density of states of the VDW structure, which is almost identical to bulk ScGaN. The inset shows the total charge density of the VDW structure.

DFT calculations further confirm the VDW structure, validating the experimentally observed increase of in-plane atomic spacing, as well as the corresponding M-M and N-N atomic distances along the [0001] direction (**Fig. 2j** and **Extended Data Fig. 3b**). While the STEM imaging performed along the [$11\bar{2}0$] direction successfully captures the in-plane lattice expansion at the domain wall, the absence of clear 8- and 4-fold rings is ascribed to a slight shift of the domain boundary within the ($11\bar{2}0$) plane (**Extended Data Fig. 4**). As illustrated in **Fig. 2k**, the nearly identical electronic structure suggests the neutral nature of the VDWs. Although similar VDW structures have been previously proposed in epitaxial films where two domains of different polarities meet, this study presents the first demonstration of electrically induced VDWs in wurtzite nitrides at atomic resolution[35, 36]. Such VDWs have been identified as efficient radiative recombination centers and could be used to fabricate lateral homojunctions[38]. The discovery of these electrically induced sub-nanometer thick VDWs thus lays the foundation for the future development of nanoscale electronic devices based on single domain walls.

## Charged horizontal domain walls

Horizontal domain walls (HDWs) are more challenging to form than their vertical counterparts due to substantial electrostatic energy arising from the extreme 180$^{o}$ polarization discontinuity. In epitaxial films, the formation of HDWs often necessitates the introduction of foreign atoms, such as silicon[39] and oxygen[40], to compensate the dangling bonds and maintain lattice symmetry. These foreign atoms substantially neutralize the polarization charges, obscuring the intrinsic electrical characteristics of the HDWs. In contrast, for ferroelectric nitrides, it is possible to create HDWs via an applied electric field without introducing foreign atoms. **Fig. 3a** shows a schematic of the HDW architecture in the ferroelectric heterostructure. A possible atomic configuration within the HDW is depicted in **Fig. 3b**. **Fig. 3c, d** showcase high magnification HAADF-STEM images and the corresponding out-of-plane distance map of a HDW observed in the polarity-reversed ScGaN capacitor. The M atom arrangement maintains the zig-zag pattern seen in wurtzite materials, highlighting the continuity of the metal sublattice in the presence of HDWs (**Extended Data Fig. 5a**). Within the central region of the HAADF image, metal-metal dimer-like patterns are again observed but exhibit contracted interatomic spacing compared to those in VDWs. Coinciding with the metal-metal dimer-like pattern formation, an out-of-plane lattice expansion is also observed (**Fig. 3d**), indicated by a sub-nanometer transition region in the M-N dumbbell angle map (**Fig. 3e, f**). Alongside the M-N dumbbells within each domain, a distinctive triangular configuration is identified in the domain wall region, with every M atom occupying a central position and symmetrically bordered by a pair of N atoms, as highlighted by the overlaid blue (N atom) and red (M atom) spheres. Compared to the N atoms in the M-N dumbbells away from the HDW, the intensities of the N atoms in the HDW are noticeably reduced, indicating decreased N atom occupation in these atomic planes. The metal atom column sites are more distinct in the dDPC image, shown in **Fig. 3g**. Subsequent atomic distance measurements based on the dDPC image are presented in **Fig. 3h**. The average N-N distance in the triangular configuration is measured to be 0.321*c*, while the average M-M distance between two different atomic layers is determined as 0.56*c* (**Fig. 3i**).

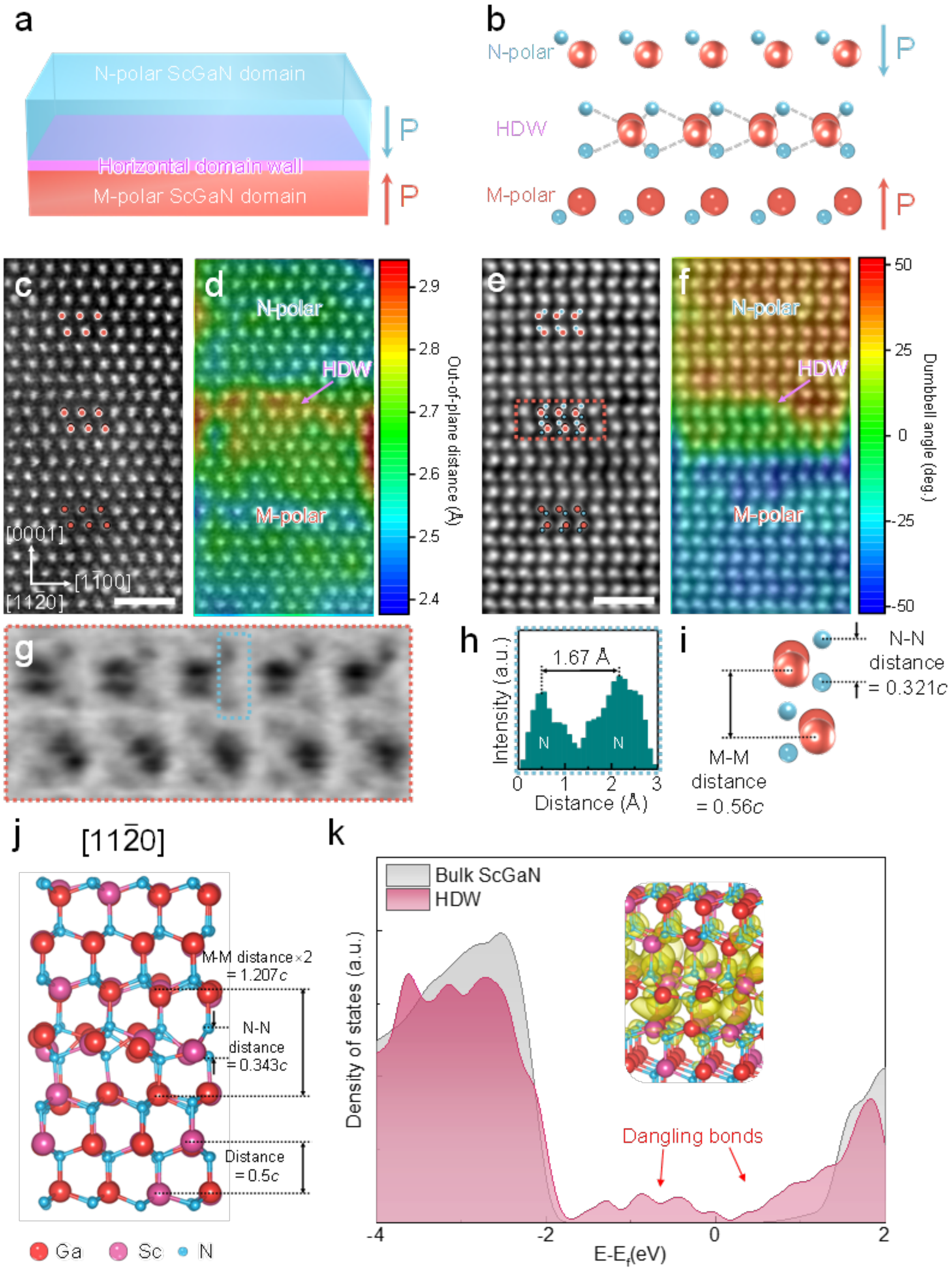

**Fig. 3. Electric-field-induced horizontal domain walls in ScGaN. a,** Schematic, and **b,** atomic model of the horizontal domain wall, with atoms represented by spheres (red for metal, blue for nitrogen). **c,** High magnification HAADF-STEM image and **d,** corresponding out-of-plane distance map of a horizontal domain wall in ScGaN. **e,** High magnification iDPC image, and **f,** corresponding dumbbell angle map for the same region shown in **c** and **d**. Scale bar for (**c-f**): 1 nm.

**g,** High resolution dDPC image of the horizontal domain wall structure. **h,** Intensity profile of the selected area in **g**. **i,** Measured average atomic distances in the triangular configurations of the horizontal domain wall structure. **j,** DFT model of the horizontal domain wall structure viewed along $[11\bar{2}0]$. The calculated average atomic distances are labeled in comparison with **i**. **k,** Comparison of the electronic density of states of the HDW structure with bulk ScGaN, demonstrating the appearance of partially occupied dangling bond states within the bandgap. The inset shows the charge density of the mid-gap dangling bond states.

DFT calculations are conducted to confirm the HDW structure. Remarkably, as depicted in the DFT model (**Fig. 3j**), despite the disorder at the metal sites, the atoms within the HDW exhibit a unique buckled 2D hexagonal phase (**Extended Data Fig. 6**). The trigonal prismatic atomic arrangement from DFT calculations aligns with the TEM observations and quantitatively matches the experimentally observed N-N spacings and out-of-plane atomic distance expansion. Along the [0001] axis, the configuration retains its inherent hexagonal symmetry, aligning seamlessly with conventional wurtzite nitrides (**Extended Data Fig. 5b**). Notably, within the domain wall, each M atom bonds with only three N atoms, creating a high density of dangling bonds (**Extended Data Fig. 7**). In **Fig. 3k**, we compare the total electronic density of states of the HDWs to that of bulk ScGaN. While the density of states of the VDWs and bulk ScGaN are nearly identical, the HDWs exhibit rich metallic-like mid-gap states due to the unpassivated dangling bonds of metal atoms in the domain wall and neighboring planes. The dangling bond nature of these metallic mid-gap states is further confirmed by the plot of their charge density (inset of **Fig. 3k**). These results suggest that the HDWs are charged domain walls and are highly electrically active.

Quantitative comparison of the charge density further reveals an intriguing interplay between the dangling bond and bound polarization discontinuity at the HDWs. As schematically shown in **Extended Data Fig. 7**, the interface exhibits two cation dangling bonds per two-dimensional unit cell, with each bond carrying an excess charge of –3*e*/4 at the HDW, reflecting the trivalent cation donating 3/4 of an electron to each of its four nitrogen neighbors (**Extended Data Fig. 8**). This generates an excess of electrons at the interface with a density of $-1.59 \times 10^{15}$ e/cm², among the highest reported for any interface. For a spontaneous polarization of ~ 1.29 C/m² for $Sc_{0.31}Ga_{0.69}N$ (**Extended Data Fig. 2**), we estimate a bound charge of $2P_{sp} = +1.61 \times 10^{15}$ e/cm² at the HDW interface. This bound positive charge is nearly perfectly compensated by the negative charge arising from the dangling bonds and thus critically contributes to the stability of the interface, potentially suppressing the formation of columnar domain patterns running through the entire film (**Extended Data Fig. 9**). Conversely, the exceptionally high polarization discontinuity at such a domain wall helps sustain this remarkably high charge density, leverageable for device applications. The reciprocal relationship between the dangling bonds and bound polarization charges, where each stabilizes the other, reveals a powerful mechanism for promoting interface stability in wurtzite ferroelectrics and opens exciting possibilities for designing novel high-performance electronics. While more accurate calculations would require taking into consideration the lattice distortion near the interface, the conclusions remain valid. We argue that the nearly complete compensation of the bound charge by the dangling bond electrons is not a coincidence, but a universal geometrical feature of tetrahedral ferroelectrics for any amount of charge transfer between cations and anions, which applies both for head-to-head and tail-to-tail antipolar domain walls (see **Methods** for a detailed discussion).

It is worth noting that both the VDWs and HDWs maintain six-fold rotational symmetry, in line with the host wurtzite material. Such symmetry is fundamental, as it necessitates the manifestation of alternate domain and domain wall projections at regular 60-degree intervals when observed from a consistent viewing angle. The atomic configuration of the HDWs suggests a vertical displacement of nitrogen atoms during polarization reversal, resulting in a 50% occupancy of the N sites in the original M-polar lattice. Conversely, the VDWs indicate a simultaneous displacement of both M and N atoms following polarity inversion. Compared to the previously postulated planar hexagonal intermediate during polarity switching in Sc-III-N alloys[41, 42], the unveiled HDW structure adopts a buckled hexagonal form that is overall non-polar along the [0001] axis. This implies that tetrahedral inversion within the HDW may occur at a lower energy threshold compared to the bulk, offering a potential pathway to mitigate the high coercive field in wurtzite ferroelectrics.

## Reconfigurable conductive domain walls in wurtzite ferroelectrics

DFT calculations shown in **Fig. 3k** suggest that the HDWs could act as conductive pathways. For ferroelectric materials with polarization directions along the out-of-plane direction, it is challenging to measure the conductivity of the HDWs directly. Here, inclined domain walls with controllable horizontal components are adopted to explore the electrical properties of the domain walls using conductive atomic force microscopy (CAFM). **Fig. 4a-c** illustrates the formation of inclined domain walls at the electrode edge in a ScGaN capacitor. Due to the asymmetric electrode configuration, the distribution of the vertical electric field near the electrode edge is not uniform. By varying the applied voltage, the inclination angle of the region reaching the coercive field can then be controlled to create inclined domain walls with different horizontal components.

To enhance the contrast in conductivity, the thickness of the ScGaN is reduced to 50 nm, yielding a positive switching voltage of ~ 20 V (**Extended Data Fig. 10**). The top electrodes are removed using dry exfoliation after poling to prevent chemical contamination and surface damage. **Fig. 4d-g** present the surface morphology, phase contrast, and conductivity maps of the electrode regions given different poling conditions (**Extended Data Fig. 10**). **Fig. 4h-k** further show the current profiles across the electrode regions extracted from the CAFM map. The appearance of conductive channels between the switched and unswitched regions is evident (**Fig. 4e, i**).

Increasing the poling voltage to 28 V leads to an increase of the domain wall inclination angle (**Fig. 4c**), reducing the horizontal component and resulting in lower conductivity compared to 21 V (**Fig. 4f, j**). More importantly, we show that applying a large negative bias can both switch back the polarity and remove the conductive channels at the electrode edge (**Fig. 4g, k**). These results point toward the horizontal components of the inclined domain walls being the principal contributors to enhanced conductivity at the electrode edge. The slight increase in the overall contrast within the poled region of **Fig. 4e, i** is attributed to the spontaneous formation of domain walls due to fluctuations in the coercive field, which vanish when the poling voltage is increased to 28 V (**Fig. 4f, j**). Our results yield the first direct observation and electrical manipulation of conductive domain walls in wurtzite semiconductors, paving the way for future investigations into their electronic and optoelectronic properties, as well as the development of functional nanodevices based on wurtzite ferroelectrics.

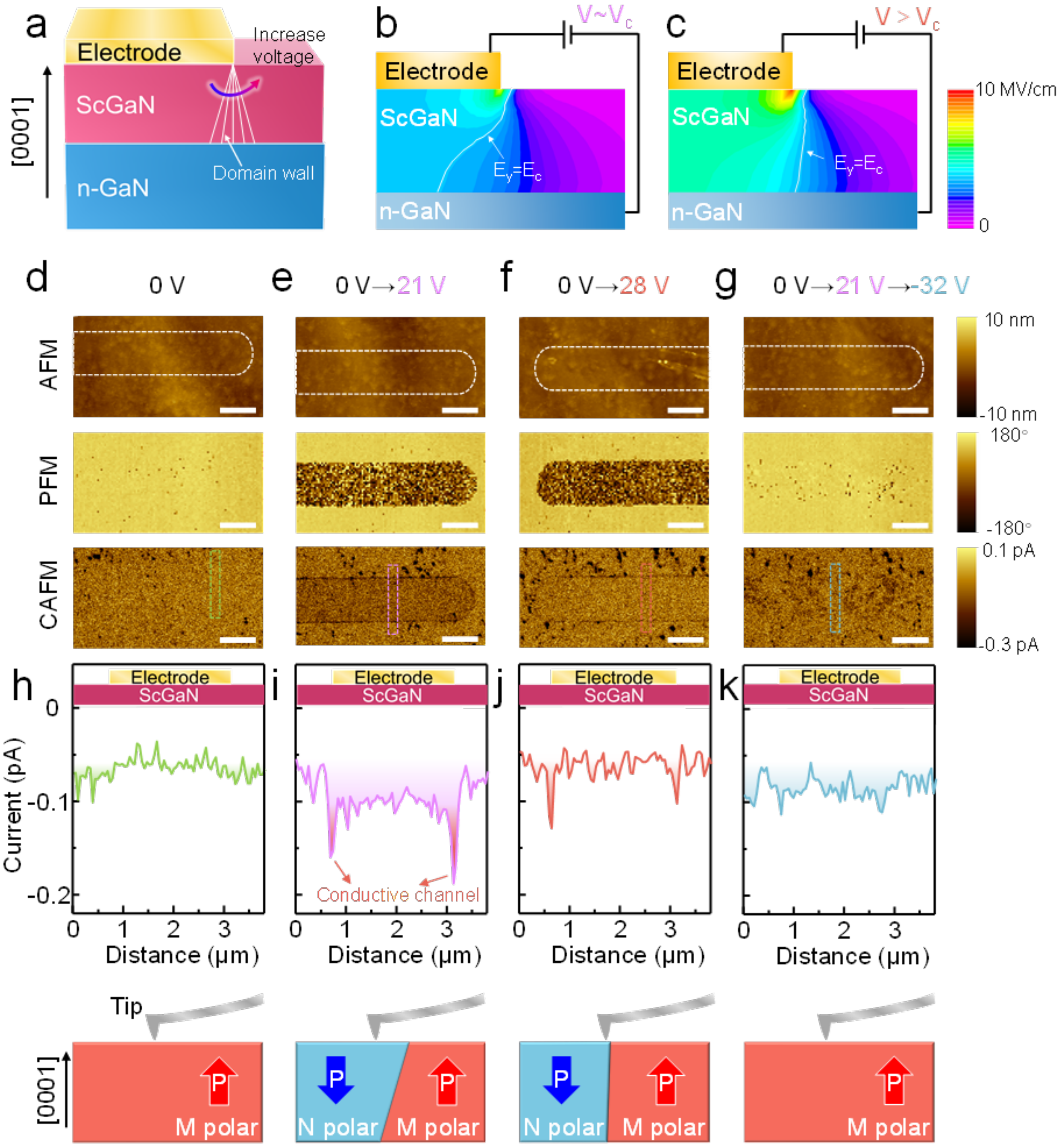

**Fig. 4. Electric field-controllable conductive domain walls in ScGaN. a,** Schematic of the inclined domain walls generated near the top electrode edge. **b, c,** Simulations of vertical electric field distributions under different voltages. **d-g,** AFM, PFM and CAFM measurements over four regions with different poling conditions. **h-k,** Current line profiles across the electrode regions, extracted from CAFM maps by averaging 20 horizontal pixels.

In conclusion, our study offers exceptional atomic-scale insights into the electric-field-induced domain walls in ScGaN, enriching our understanding of the microstructures and domain dynamics in wurtzite ferroelectrics. The agreement between our STEM observations and DFT calculations validates the atomic configuration of these domain walls, marking a key advancement towards optimizing the ferroelectric properties of wurtzite materials for practical applications. The novel buckled 2D hexagonal phase with abundant dangling bonds further fosters the demonstration of reconfigurable conductive domain walls. With their atomic-scale thickness and readily tunable properties via external electric fields, these domain walls hold significant possibilities in the design of next-generation nanoscale devices. Furthermore, the universal charge compensation mechanism significantly augments our understanding of the key factors that underpin the stability of wurtzite ferroelectric devices. The depth of understanding we have achieved herein provides critical insights for the design and development of future wurtzite ferroelectric materials and their applications in various technologies.

**Main References**

1. Kim, K. H., Karpov, I., Olsson III, R. H. & Jariwala, D. Wurtzite and fluorite ferroelectric materials for electronic memory. *Nat. Nanotechnol.* **18**, 422-441 (2023).
2. Wang, P. *et al.* Dawn of nitride ferroelectric semiconductors: from materials to devices. *Semicond. Sci. Technol.* **38**, 043002 (2023).
3. Mikolajick, T. *et al.* Next generation ferroelectric materials for semiconductor process integration and their applications. *J. Appl. Phys.* **129**, 100901 (2021).
4. Fichtner, S., Wolff, N., Lofink, F., Kienle, L. & Wagner, B. AlScN: A III-V semiconductor based ferroelectric. *J. Appl. Phys.* **125**, 114103 (2019).
5. Hardy, M. T. *et al.* Epitaxial ScAlN grown by molecular beam epitaxy on GaN and SiC

substrates. *Appl. Phys. Lett.* **110**, 162104 (2017).

6. Calderon, S. *et al.* Atomic-scale polarization switching in wurtzite ferroelectrics. *Science* **380**, 1034-1038 (2023).
7. Zhu, W. *et al.* Wake-up in $Al_{1-x}B_xN$ ferroelectric films. *Adv. Electron. Mater.* **8**, 2100931 (2022).
8. Wang, D., Wang, P., Wang, B. Y. & Mi, Z. Fully epitaxial ferroelectric ScGaN grown on GaN by molecular beam epitaxy. *Appl. Phys. Lett.* **119**, 111902 (2021).
9. Wang, D. *et al.* Ferroelectric YAlN grown by molecular beam epitaxy. *Appl. Phys. Lett.* **123**, 033504 (2023).
10. Schönweger, G. *et al*. Ultrathin $Al_{1-x}Sc_xN$ for low-voltage-driven ferroelectric-based devices. *Phys. Status Solidi RRL* **17**, 2200312 (2023).
11. Zheng, J. X. *et al*. Ferroelectric behavior of sputter deposited $Al_{0.72}Sc_{0.28}N$ approaching 5 nm thickness. *Appl. Phys. Lett.* **122**, 222901 (2023).
12. Pradhan, D. K. *et al.* Scalable and stable ferroelectric non-volatile memory at > 500 ℃. *arXiv preprint arXiv:2309.04555* (2023).
13. Wang, D. *et al.* An epitaxial ferroelectric ScAlN/GaN heterostructure memory. *Adv. Electron. Mater.* **8**, 2200005 (2022).
14. Islam, M. R. *et al*. On the exceptional temperature stability of ferroelectric $Al_{1-x}Sc_xN$ thin films. *Appl. Phys. Lett.* **118**, 232905 (2021).
15. Guido, R. *et al*. Thermal stability of the ferroelectric properties in 100 nm-thick $Al_{0.72}Sc_{0.28}N$. *ACS Appl. Mater. Interfaces* **15**, 7030-7043 (2023).
16. Lin, B. *et al.* A high Q value ScAlN/AlN-based SAW resonator for load sensing. *IEEE Trans. Electron Devices* **68**, 5192-5197 (2021).
17. Wang, D. *et al.* Ultrathin nitride ferroic memory with large ON/OFF ratios for analog in-memory computing. *Adv. Mater.* **35**, 2210628 (2023).

18. Liu, X. *et al*. Reconfigurable compute-in-memory on field-programmable ferroelectric diodes. *Nano Lett.* **22**, 7690-7698 (2022).

19. Liu, X. *et al*. Post-CMOS compatible aluminum scandium nitride/2D channel ferroelectric field-effect-transistor memory. *Nano Lett.* **21**, 3753-3761 (2021).

20. Kim, K. H. *et al.* Scalable CMOS back-end-of-line-compatible AlScN/two-dimensional channel ferroelectric field-effect transistors. *Nat. Nanotechnol.* **18**, 1044-1050 (2023).

21. Wen, Z. & Wu, D. Ferroelectric tunnel junctions: modulations on the potential barrier. *Adv. Mater.* **32**, 1904123 (2020).

22. Schönweger, G. *et al*. In-grain ferroelectric switching in sub-5 nm thin $Al_{0.74}Sc_{0.26}N$ films at 1 V. *Adv. Sci.* **10**, 2302296 (2023).

23. Wolff, N. *et al*. Atomic scale confirmation of ferroelectric polarization inversion in wurtzite-type AlScN. *J. Appl. Phys.* **129**, 034103 (2021).

24. Schönweger, G. *et al*. From fully strained to relaxed: epitaxial ferroelectric $Al_{1-x}Sc_xN$ for III-N technology. *Adv. Funct. Mater.* **32**, 2109632 (2022).

25. Yazawa, K. *et al.* Anomalously abrupt switching of wurtzite-structured ferroelectrics: simultaneous non-linear nucleation and growth model. *Mater. Horiz.* **10**, 2936-2944 (2023).

26. Sharma, P., Moise, T. S., Colombo, L. & Seidel, J. Roadmap for ferroelectric domain wall nanoelectronics. *Adv. Funct. Mater.* **32**, 2110263 (2022).

27. Meier, D. & Selbach, S. M. Ferroelectric domain walls for nanotechnology. *Nat. Rev. Mater.* **7**, 157-173 (2022).

28. Yang, W. *et al.* Nonvolatile ferroelectric-domain-wall memory embedded in a complex topological domain structure. *Adv. Mater.* **34**, 2107711 (2022).

29. Wang, P. *et al.* Fully epitaxial ferroelectric ScAlN grown by molecular beam epitaxy. *Appl. Phys. Lett.* **118**, 223504 (2021).

30. Wang, P. *et al.* Quaternary alloy ScAlGaN: a promising strategy to improve the quality of ScAlN. *Appl. Phys. Lett.* **120**, 012104 (2022).

31. Wang, P. *et al.* Oxygen defect dominated photoluminescence emission of ScAlN grown by molecular beam epitaxy. *Appl. Phys. Lett.* **118**, 032102 (2021).

32. Wang, Y. *et al.* Ferroelectric dead layer driven by a polar interface. *Phys. Rev. B* **82**, 094114 (2010).

33. Wurfel, P. & Batra, I. P. Depolarization-field-induced instability in thin ferroelectric films-experiment and theory. *Phys. Rev. B* **8**, 5126-5133 (1973).

34. Catalan, G., Seidel, J., Ramesh, R. & Scott, J. F. Domain wall nanoelectronics. *Rev. Mod. Phys.* **84**, 119-156 (2012).

35. Romano, L. T., Northrup, J. E. & Okeefe, M. A. Inversion domains in GaN grown on sapphire. *Appl. Phys. Lett.* **69**, 2394-2396 (1996).

36. Northrup, J. E., Neugebauer, J. & Romano, L. T. Inversion domain and stacking mismatch boundaries in GaN. *Phys. Rev. Lett.* **77**, 103-106 (1996).

37. Nord, M., Vullum, P. E., MacLaren, I., Tybell, T. & Holmestad, R. Atomap: a new software tool for the automated analysis of atomic resolution images using two-dimensional Gaussian fitting. *Adv. Struct. Chem. Imaging* **3**, 9 (2017).

38. Stutzmann, M. *et al*., Playing with polarity. *Phys. Status Solidi B* **228**, 505-512 (2001).

39. Wang, P. *et al.*, Interfacial modulated lattice-polarity-controlled epitaxy of III-nitride heterostructures on Si(111). *ACS Appl. Mater. Interfaces* **14**, 15747-15755 (2022).

40. Liu, F. *et al.*, Lattice polarity manipulation of quasi-vdW epitaxial GaN films on graphene through interface atomic configuration. *Adv. Mater.* **34**, 2106814 (2022).

41. Liu, Z., Wang, X., Ma, X., Yang, Y. & Wu, D. Doping effects on the ferroelectric properties of wurtzite nitrides. *Appl. Phys. Lett.* **122**, 122901 (2023).

42. Zhang, S., Holec, D., Fu, W. Y., Humphreys, C. J. & Moram, M. A. Tunable optoelectronic and ferroelectric properties in Sc-based III-nitrides. *J. Appl. Phys.* **114**, 133510 (2013).

# Methods

## Sample preparation

The ScGaN/GaN heterostructure was grown on 2-inch n-GaN/sapphire templates using a Veeco GENxplor MBE system equipped with a radio frequency (RF) plasma-assisted nitrogen source for supplying active nitrogen ($N^*$, 99.9999%), dual filament SUMO cells for providing gallium (Ga, 99.99999%), and a high-temperature Knudsen effusion cell for providing scandium (Sc, 99.999%). The GaN layer was highly Si-doped with a carrier concentration of ~ $2\times10^{19}$ $cm^{-3}$, serving as the bottom electrode. The ScGaN layer was grown under N-rich conditions with a thickness of ~100 nm. A nominal Sc content of 0.31 was measued by energy dispersive X-ray spectroscopy (EDX) analysis performed in a scanning electron microscope (SEM, Hitachi SU8000) using a seperate sample grown on AlN substrate under the same growth conditons.

## Microstructure imaging and determination of lattice parameters

TEM was carried out using a Thermo Fisher Scientific Spectra 300 probe-corrected STEM equipped with a Dual-X EDS system and operated at 300 kV. High-angle annular dark-field (HAADF) images were collected in a range of 62–200 mrad with a convergence angle of 22 mrad. A non-linear filter was adopted to reduce noise[43]. The cross-sectional TEM specimen was prepared using an in-situ focused ion beam (FIB) lift-out method employing a Thermo Fisher Helios 600 Xe plasma FIB/SEM system. Initially, the sample was cut using a high current at 30 kV for the lift-out process. Subsequently, to remove the damage induced by the high energy beam, a final thinning process was performed using a 5 kV beam at 10 pA. The specimen's thickness was carefully monitored by observing the secondary electron image in the SEM at 5 kV. This process continued until the sample appeared bright in the image, indicating that it had been thinned to an approximate thickness of ~100 nm. The central position associated with each atomic column was localized by using a two-dimensional Gaussian fitting procedure[37]. The interplanar spacings used in the in-plane and out-of-plane distance mapping are the distances between equivalent atomic positions across successive planes.

**DFT calculations**

Density functional theory (DFT) calculations were performed using the Vienna ab initio simulation package (VASP )[44-47]. The generalized gradient approximation for the exchange-correlation functional[48] and projector augmented wave (PAW) potentials was employed. Structural optimization calculations were performed for ScGaN alloys with a 31.25% Sc composition. The disordered environment of the alloy is simulated by randomly distributing the Sc atoms over the cation sites. The vertical domain walls were modelled by constructing a 4×4×2 supercell (128 atoms) that includes one vertical domain wall. To model the horizontal domain wall, a 224-atom structure consisting of two polar wurtzite slabs with opposite polarities surrounded by 12.4 Å of vacuum was constructed. The metal atoms occupy a common sublattice for both polarity regions, while the interface nitrogen atoms are randomly assigned to belong either to the N-polar or the metal-polar domain. In this work, the plane-wave cutoff energy was set at 500 eV and integrations over the Brillouin zone using a 2×2×1 shifted k-point grid was performed. The structures were subjected to relaxation until the energies converged to $10^{-6}$ eV and the forces converged to 0.005 eV/Å. The dangling bonds at the N-terminated outer surfaces were passivated with pseudohydrogen atoms with ¾ charge. Both domain wall configurations have their in-plane lattice constants constrained to that of GaN in order to simulate the epitaxial growth of the ScGaN/GaN heterostructures. The density of states of all the structures and the band-decomposed charge density for the dangling bonds of the horizontal domain wall structure were further calculated. For visualization, the Visualization for Electronic and Structural Analysis (VESTA) software was employed[49].

## Electrical characterization

Ti/Au (20 nm/120 nm) top electrodes were deposited in an electron-beam evaporator and defined via a photolithography and lift-off process. For piezo-response force microscopy (PFM) measurements, to remove the metal electrodes, Ti/Al (20 nm/120 nm) instead of Ti/Au metal stacks were used. The ferroelectric properties, including P-E and J-E loops, pulse transients, and Positive-Up-Negative-Down (PUND) measurements were conducted using a Radiant Precision Premier II ferroelectric tester driven from the top electrode. A Keithley 2400 SMU was used to set the polarity before TEM characterization.

## Characterization of conductive domain walls

50 nm ScGaN was grown under the same conditions as above to explore the conductive domain walls. The thickness was reduced to 50 nm to enhance the conductivity contrast. To avoid possible chemical contamination or etching damage to the surface, a 120 nm Au metal stack was patterned via e-beam deposition and lithography and exfoliated after electrical poling using a dry transfer method. Triangular pulses with a period of 1s were used to pole the devices. A switching voltage of 20 V was extracted, which was used as reference for selecting the poling voltages in **Fig. 4**. The PFM and CAFM measurements were carried out using a Bruker ICON AFM. For PFM, the tip frequency was 15 kHz and $V_{AC}$ was 10 V applied to the sample. For CAFM, SCM-PIT conductive tips were used under contact mode-based tunneling AFM (TUNA). During measurements, the tip frequency was 2.5 kHz and sample bias was -10 V. The electric field distribution was modeled using Silvaco TCAD.

**Interplay between polarization discontinuity and dangling bonds**

By referencing the polarization to the planar-hexagonal BN-like structure, the displacement of the positive and negative charge of the dipoles (nominally $3e$ for ScGaN) is the distance from the base to the centroid of the tetrahedron, $(1/2-u)c$, with $u$ approximately close to the ideal value, $3/8$, for the regular tetrahedron (**Extended Data Fig. 8**). This yields a dipole moment of

$$\vec{p}=3e\left(\frac{1}{2}-u\right)c\hat{z}$$

and a polarization per wurtzite unit cell (containing two such dipoles) of

$$\vec{P}=2\vec{p}/V_{cell}=6e\left(\frac{1}{2}-u\right)\hat{z}/\left(a^2\frac{\sqrt{3}}{2}\right)$$

Thus, the surface-bound charge at the antipolar domain is

$$\sigma_b=2\left|\vec{P}\right|\approx\frac{3}{2}e\bigg/\left(a^2\frac{\sqrt{3}}{2}\right)$$

At the same time, each cation dangling bond at the interface carriers an excess charge of $-\frac{3}{4}e$, and thus the excess electron density arising from the two dangling bonds per unit cell for the HDW structure at the interface is

$$\sigma_e=-2\frac{3}{4}e/\left(a^2\frac{\sqrt{3}}{2}\right)$$

which is equal and opposite to the surface bound charge. Note that this cancellation is equally true for any amount of charge transfer between cations and anions, even fractional, and it applies both for head-to-head and tail-to-tail antipolar domain walls. Therefore, the dangling bonds at the HDW interface structure provide a universal mechanism that compensates the surface-bound charge and stabilizes antipolar domain walls in tetrahedral ferroelectrics.

**Data and materials availability**

The data and material generated and analysed during the current study are available from the corresponding authors on reasonable request.

43. Du, H., A nonlinear filtering algorithm for denoising HR(S)TEM micrographs. *Ultramicroscopy* **151**, 62-67 (2015).
44. Kresse, G. & Hafner, J. Ab initio molecular dynamics for liquid metals. *Phys. Rev. B* **47**, 558 (1993).
45. Kresse, G. & Furthmüller, J. Efficiency of ab-initio total energy calculations for metals and semiconductors using a plane-wave basis set. *Comput. Mater. Sci.* **6**, 15-50 (1996).
46. Kresse, G. & Furthmüller, J. Efficient iterative schemes for ab initio total-energy calculations using a plane-wave basis set. *Phys. Rev. B* **54**, 11169 (1996).
47. Kresse, G. & Joubert, D. From ultrasoft pseudopotentials to the projector augmented-wave method. *Phys. Rev. B* **59**, 1758 (1999).
48. Perdew, J. P., Burke, K. and Ernzerhof, M. Generalized gradient approximation made simple. *Phys. Rev. Lett.* **77**, 3865-3868 (1996).
49. Momma, K. and Izumi, F. VESTA 3 for three-dimensional visualization of crystal, volumetric and morphology data. *J. Appl. Crystallogr.* **44**, 1272-1276 (2011).

## Acknowledgments

This work is supported by the National Science Foundation under Grant Nos. 2235377 (experimental studies) and 2329109 (theoretical calculations). The authors also acknowledge the technical support from the Lurie Nanofabrication Facility (LNF) and the Michigan Center for Materials Characterization [$(MC)^2$] at the University of Michigan. Computational resources are provided by the National Energy Research Scientific Computing Center, which is supported by the Office of Science of the U.S. Department of Energy under Contract No. DE-AC02-05CH11231.

## Author contributions

D.W., D.H.W. and M.M. contributed equally to this work. Z.M., D.W. and D.H.W. conceived the idea. D.W. and D.H.W. performed the MBE growth, device fabrication and ferroelectric measurements. D.W., D.H.W., S.Y., M.T.H., J.N.L. and Y.P.W. collected data and performed the material characterization. D.W., D.H.W., and T.M. carried out the TEM characterization and analysis. M.M., Y.J.L., and E.K. conducted and discussed the theoretical calculations. D.W., D.H.W., T.M., M.M., Y.J.L., E.K. and Z.M. discussed the data analysis. D.W., D.H.W., and Z.M. wrote the initial draft of the manuscript, which was then revised by other co-authors. All authors have given approval for the final version of the manuscript.

## Competing interests

The authors declare that they have no competing interests.

## Extended data/Figures

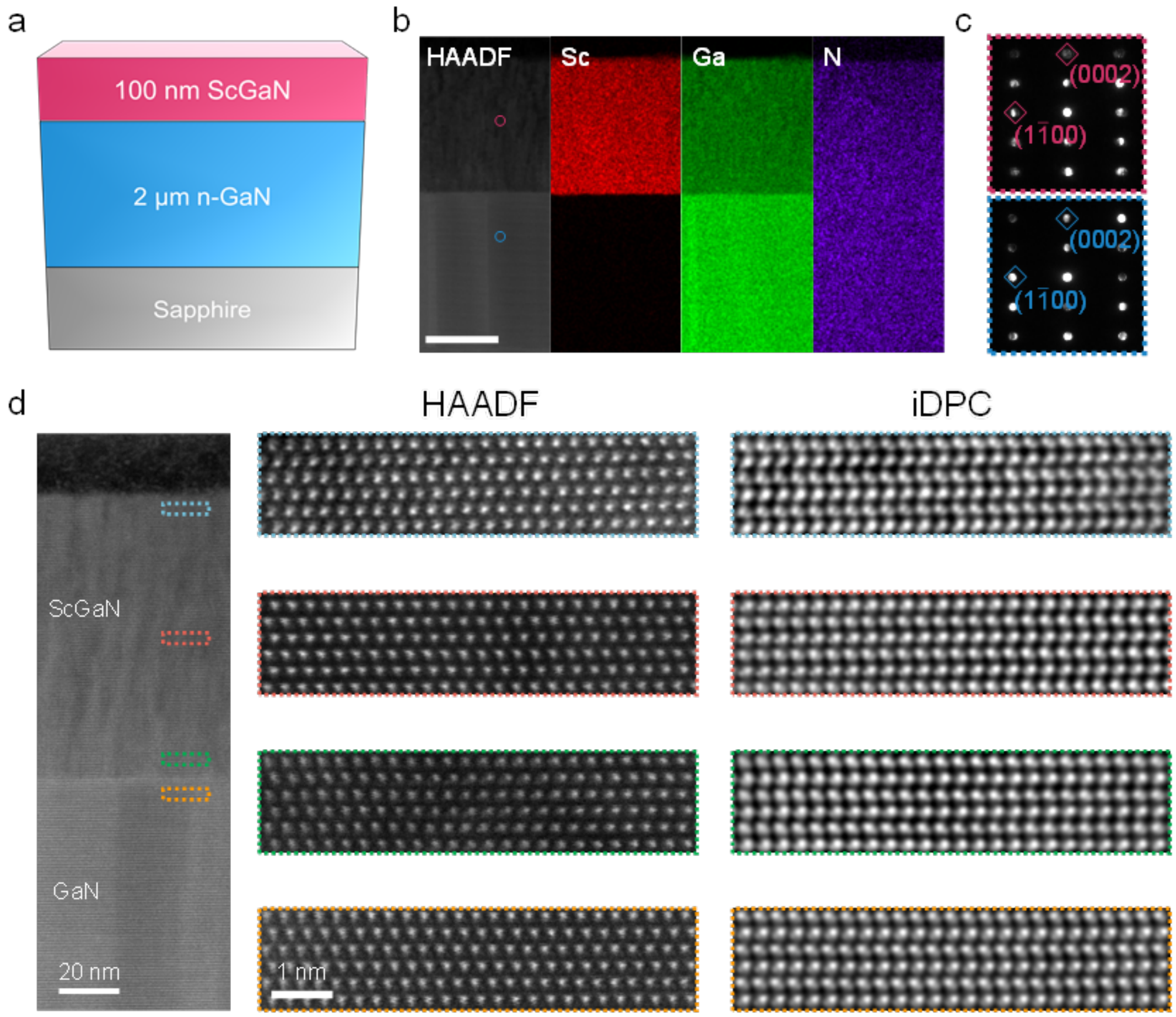

**Extended Data Fig. 1. Microstructure of as-grown ScGaN on GaN template. a,** Schematic of the sample structure. **b,** Cross-sectional HAADF-STEM image and corresponding elemental map, indicating uniform elemental distribution. The wurtzite lattice group of the ScGaN layer is further evidenced by the nanobeam electron diffraction (NBED) patterns in **(c)**. Scale bar: 50 nm. **d**, High resolution STEM images of as-grown ScGaN on GaN. Magnified images with HAADF and iDPC contrast are taken from different regions in the heterostructure as indicated in the low magnification STEM image on the left, showing uniform metal-polarity in the ScGaN film before ferroelectric switching. This further supports that the N-polar regions in switched capacitors result from electric-field-induced polarity switching rather than being attributable to localized defects inherent to the material.

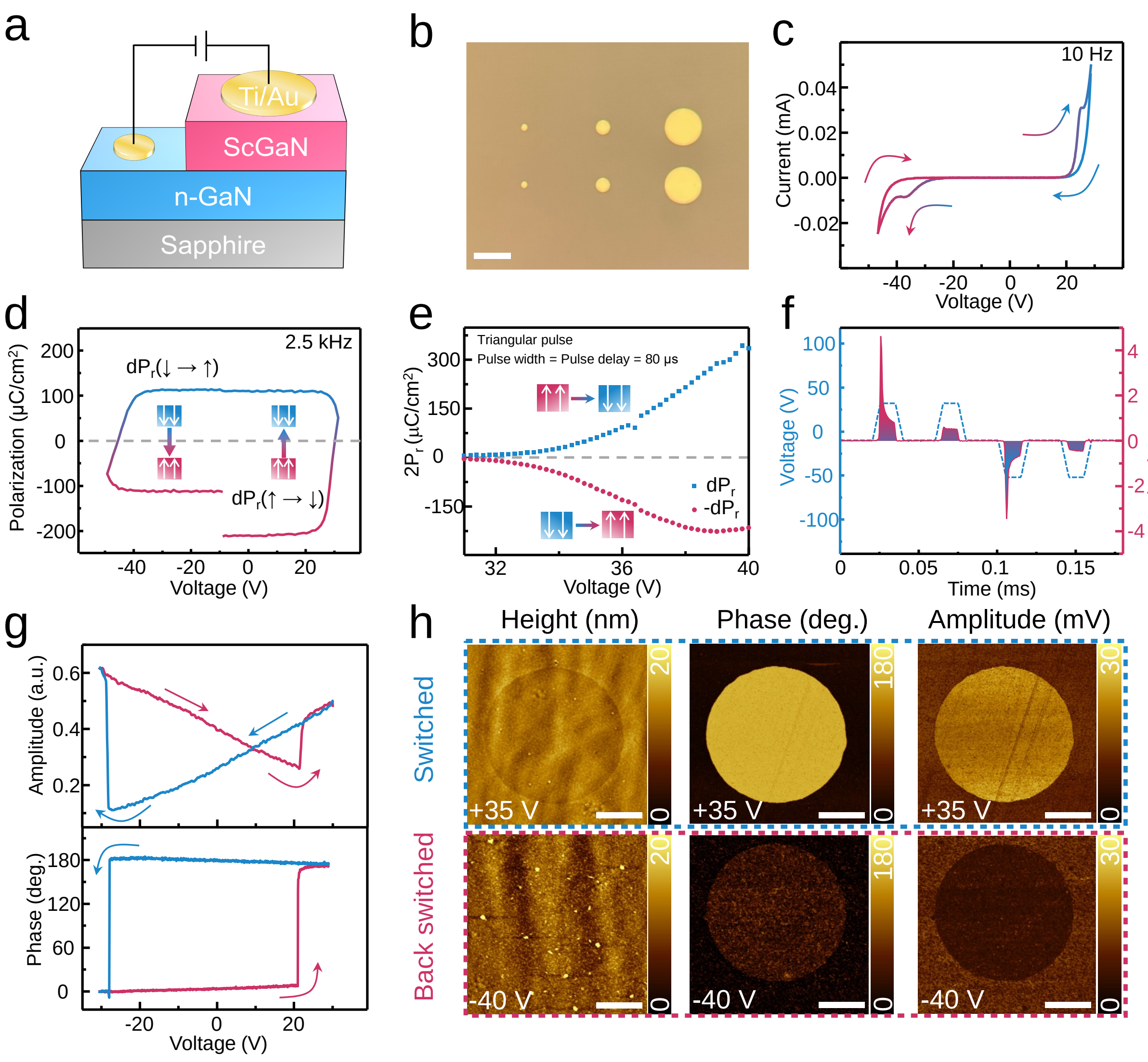

**Extended Data Fig. 2. Ferroelectric switching in Metal/ScGaN/n-GaN capacitors. a,** Schematic of the device structure. **b,** Optical microscope image of the fabricated capacitors (Scale bar: 50 μm). Highly doped n-GaN serves as the bottom electrode, with a metal stack functioning as the top electrode. **c,** J-V and **d,** P-V loops of the capacitor using triangular voltage profiles at 10 Hz and 2.5 kHz, respectively, at room temperature. Displacement current peaks can be observed in both directions. Non-switching currents have been subtracted to better show the hysteresis loop in (**d**). **e,** PUND measurement results using triangular voltage pulses with a pulse width of 80 μs, suggesting a remanent polarization of ~ 129 $\mu C/cm^2$ (ascertained by averaging the $P_r$ value at 39

V). Arrows within the figures highlight the polarization direction of the ScGaN layer before and after electrical switching. **f,** Exemplary pulse transients using trapezoidal voltage pulses to better show the displacement current. Prior to executing polarization measurements, multiple bipolar poling cycles are performed to fully wake up the material[8]. **g,** PFM amplitude and phase hysteresis loops of ScGaN. A clear butterfly-shape amplitude diagram and a box-shape phase diagram with approximately 180° separations can be observed, showing clear ferroelectric polarization switching under external biasing. **h,** PFM phase and amplitude maps after poling the capacitor (circular regions) with different voltages. The electrodes are removed using HF before PFM measurements (Scale bar: 10 μm). During measurements, the tip frequency is 15 kHz and $V_{AC}$ is 10 V applied to the tip.

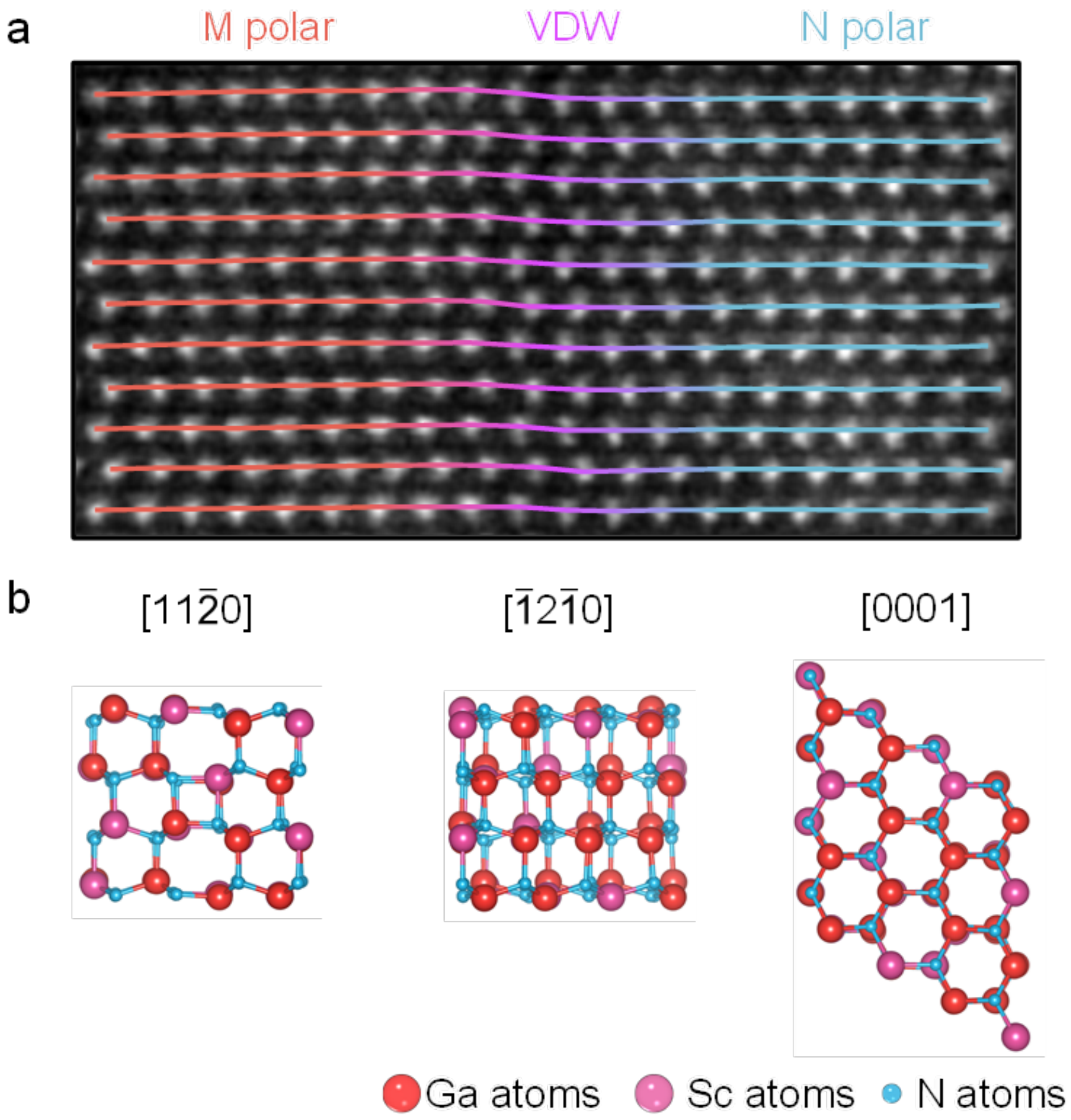

**Extended Data Fig. 3. Metal plane displacement across VDWs and the proposed VDW structure viewed along different projections. a,** High magnification HAADF-STEM image of the electric-field-induced VDW in ScGaN. The auxiliary lines within the figure traces the center of each metal atom column. The position of the center is obtained from 2D Gaussian fitting. A shift in the metal planes while crossing a VDW is evident. **b,** the proposed vertical domain wall structure viewed along different projections. The structure maintains the six-fold rotational symmetry of the host wurtzite material and satisfies the symmetry requirement on the manifestation of alternate domain and domain wall projections at regular 60-degree intervals when observed from a consistent viewing angle.

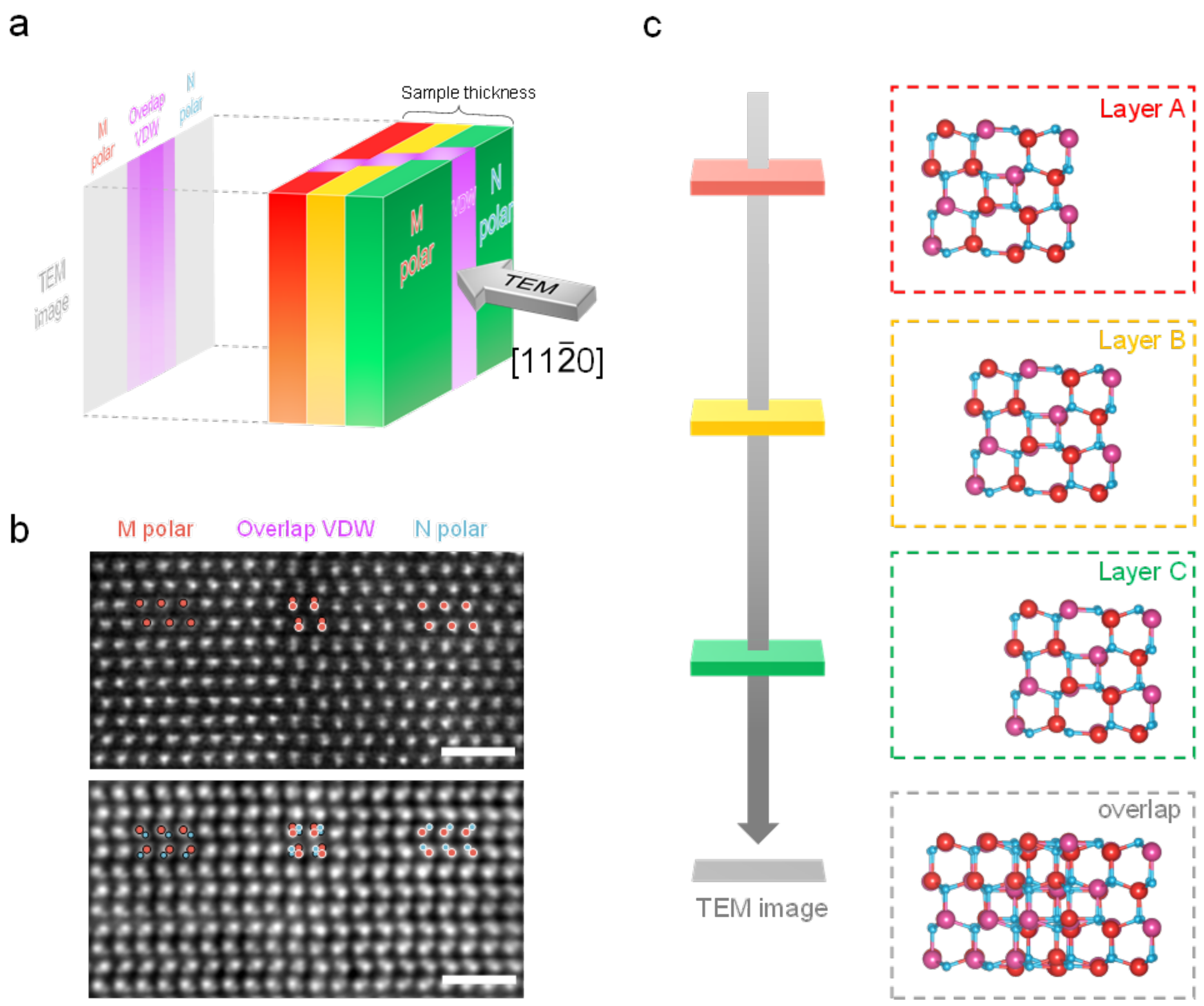

**Extended Data Fig. 4. Schematic illustration depicting the formation of dimer-like atomic configurations in VDWs.** **a,** Simplified schematic of acquiring STEM images and **b,** obtained STEM results. **c,** Conceptual illustration explaining the origin of overlapping patterns in vertical domain walls (VDWs). Despite STEM imaging being conducted along the [11$\bar{2}$0] direction, the expected 8- and 4-fold ring configuration is not always discernible. This can be attributed to a slight shift of the domain wall relative to the [11$\bar{2}$0] viewing direction.

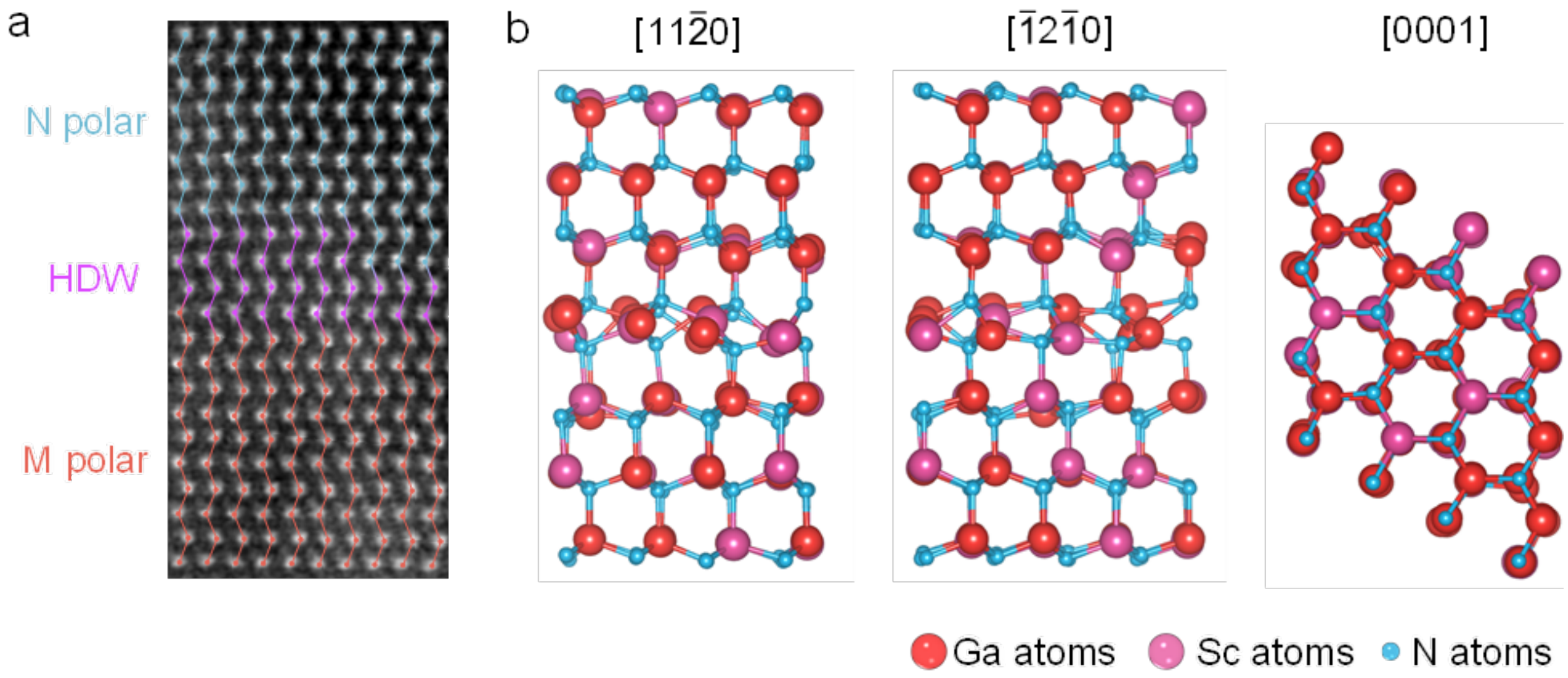

**Extended Data Fig. 5. Zig-zag feature of the metal atoms across HDWs and the proposed HDW structure viewed along different projections. a,** High magnification HAADF-STEM images of a horizontal domain wall in ScGaN. The auxiliary lines within the figure traces the center of each metal atom column. The position of the center is obtained from 2D Gaussian fitting. **b,** The proposed horizontal domain wall structure viewed along different projections. The structure maintains the zig-zag fashion and six-fold rotational symmetry of the host wurtzite material and satisfies the symmetry requirement on the manifestation of alternate domain and domain wall projections at regular 60-degree intervals when observed from a consistent viewing angle.

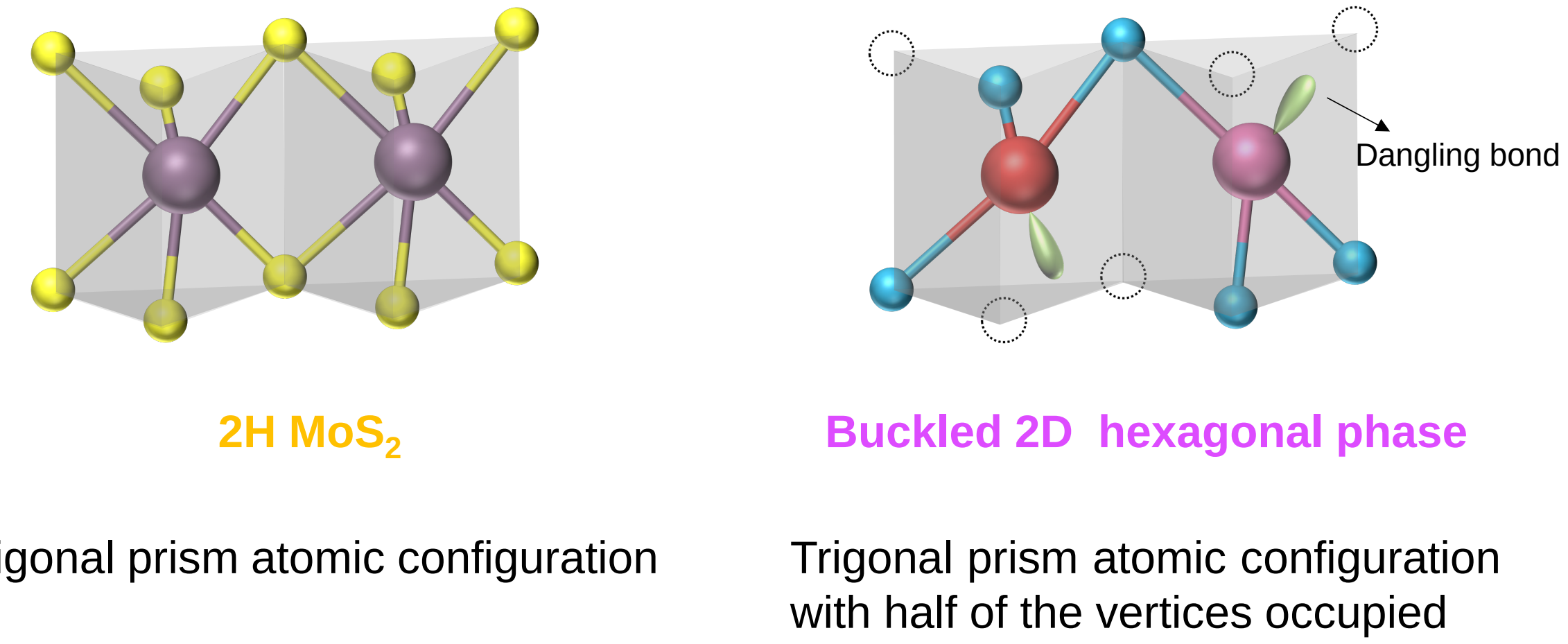

**Extended Data Fig. 6. Structural comparison of 2H $MoS_2$ and the proposed HDW model**. Both 2H $MoS_2$ and HDWs in ScGaN exhibit a trigonal prism atomic configuration. However, a distinctive feature of ScGaN HDWs is that only half of the vertices are occupied, coupled with the presence of one dangling bond at each metal site.

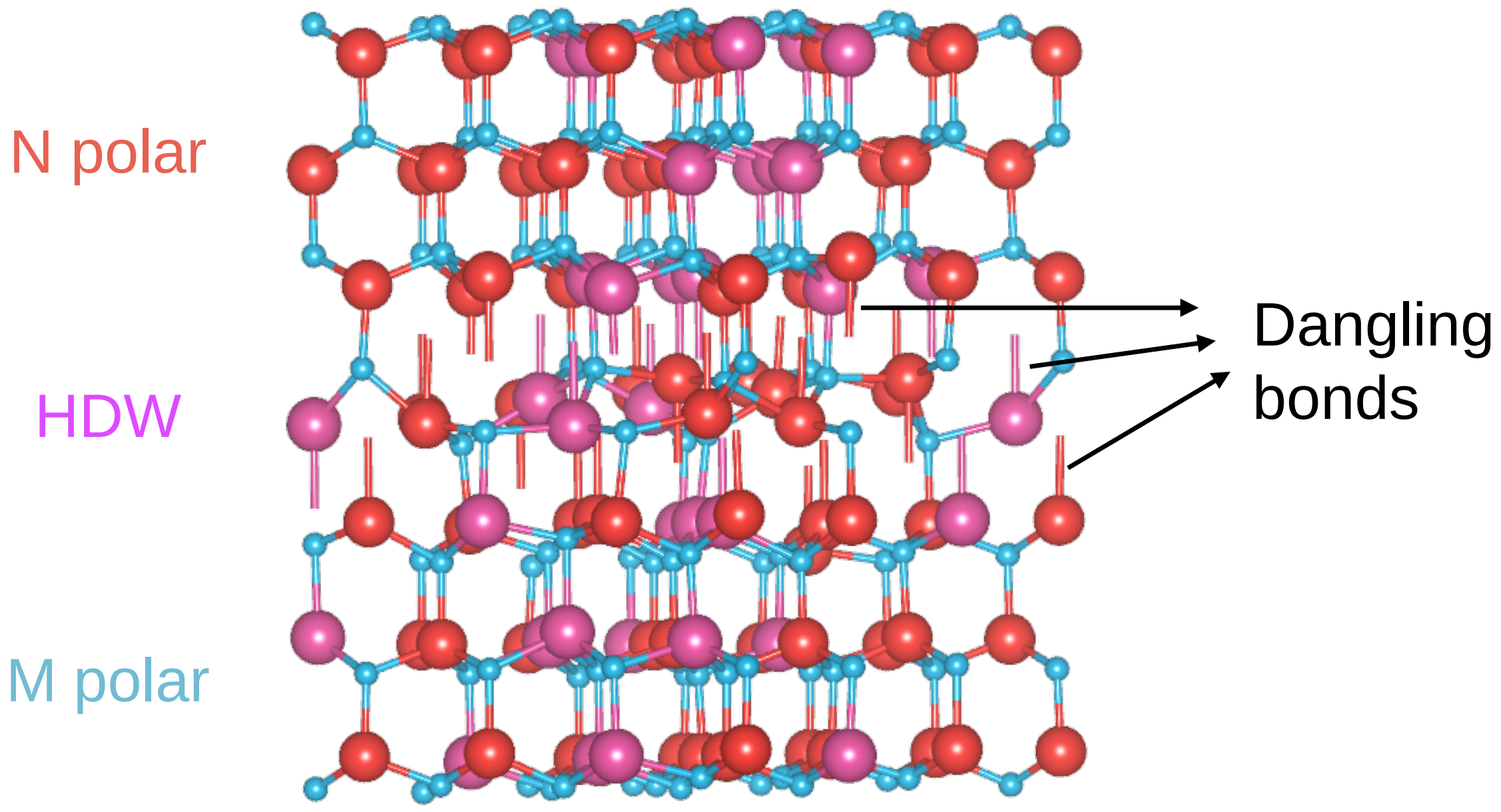

**Extended Data Fig. 7. Calculated structural model of the HDW structure with dangling bonds**. Note that these representations primarily showcase the presence of dangling bonds on the atoms. The specific orientations of these bonds, as depicted, are illustrative in nature and do not imply any physical significance.

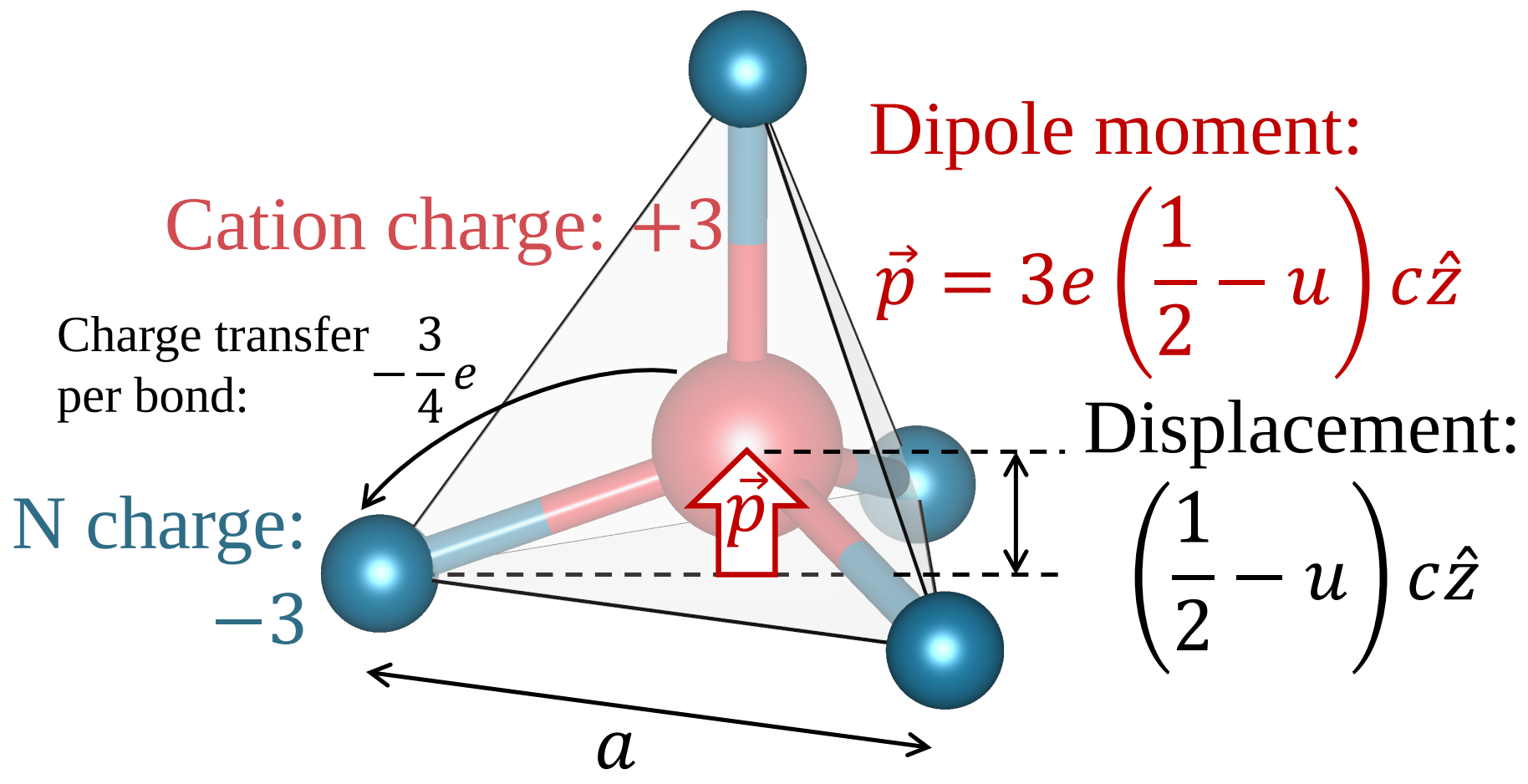

**Extended Data Fig. 8. Polarization estimation in tetrahedral ferroelectrics.** Blue atoms are anions and the red atom represents the cation. The estimated charge transfer, atom displacement, and resultant polarization, referencing the planar-hexagonal BN-like structure, have been indicated.

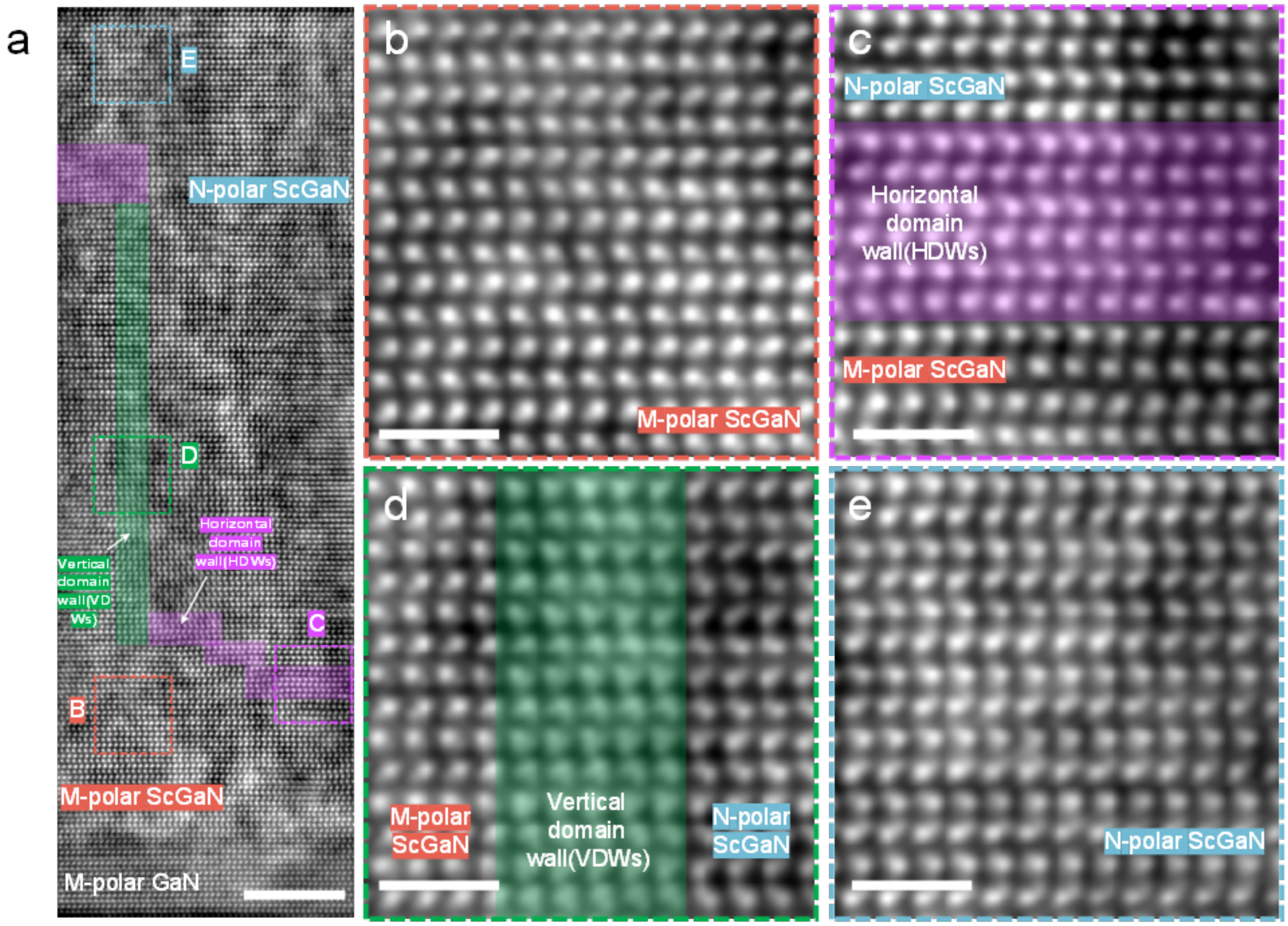

**Extended Data Fig. 9. Larger scale TEM image showcasing the co-existence of VDWs and HDWs. a,** High resolution iDPC-STEM image near the ScGaN/GaN interface. (**b** to **e**) Magnified views of the regions B, C, D, E from (**a**). Area D reveals the presence of vertical domain walls, whereas areas B and E display exclusively M-polarity and N-polarity, respectively. This observation suggests that the vertical domain walls have a limited extent and supports our interpretation that these walls are localized manifestations of the undulating interface. The scale bars in A and B-E are 5 nm and 1 nm, respectively.

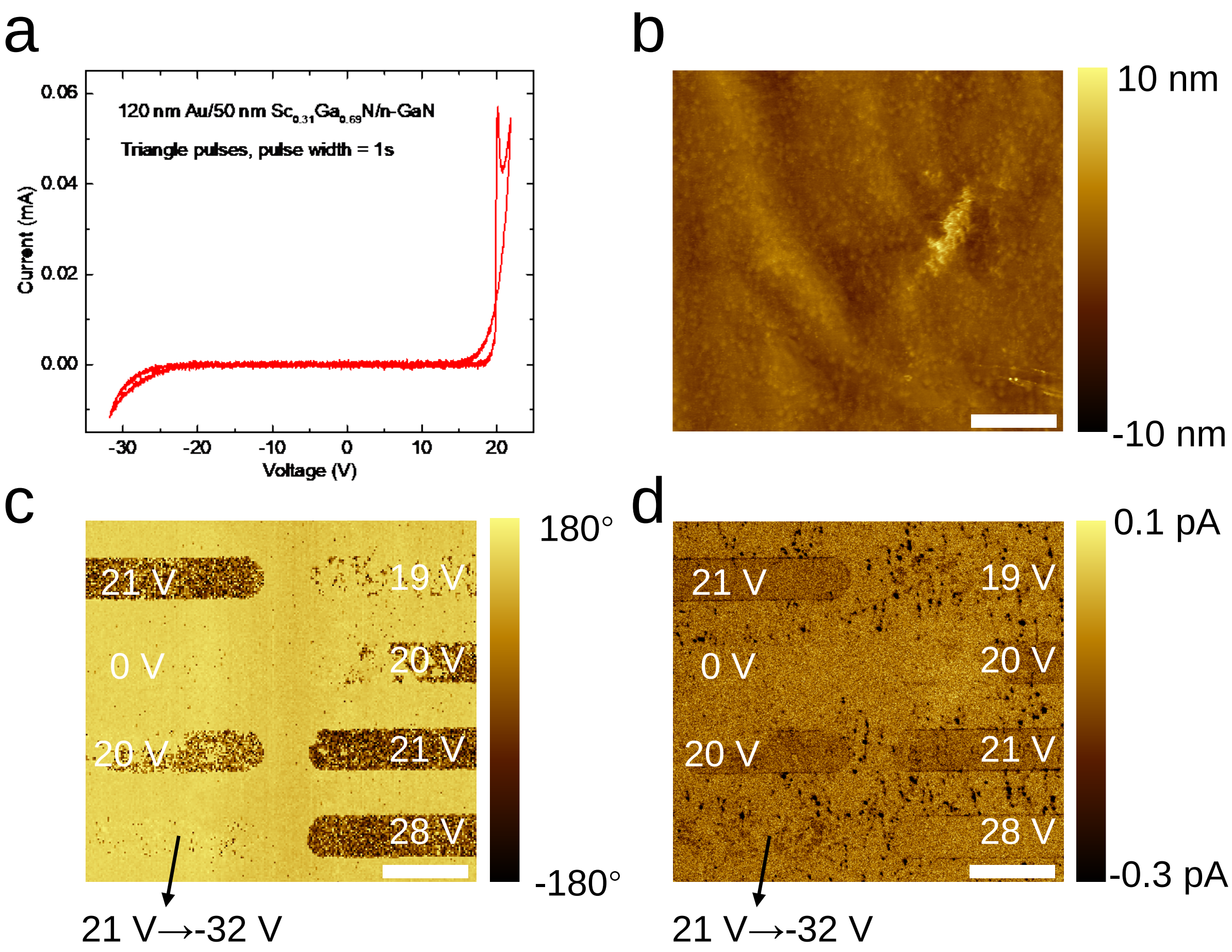

**Extended Data Fig. 10. Poling voltage dependent conductive domain walls in ScGaN. a,** Displacement current measurement results for 50 nm ScGaN capacitors. (**b** to **d**) AFM, PFM and CAFM measurement results, showcasing the formation and erasure of voltage dependent conductive domain walls near the edge of electrodes. Scale bar: 2 μm.